\documentstyle[12pt]{article}
\newcommand{\newsection}{
\setcounter{equation}{0}
\section}
\newcommand{\rf}[1]{(\ref{#1})}
\newcommand{\bea}{\begin{eqnarray}}
\newcommand{\eea}{\end{eqnarray}}
\newcommand{\g}{\gamma}
\renewcommand{\l}{\lambda}
\renewcommand{\b}{\beta}
\renewcommand{\a}{\alpha}
\newcommand{\n}{\nu}
\newcommand{\m}{\mu}

\newcommand{\sg}{\sigma}
\newcommand{\kp}{\kappa}

\newcommand{\Dl}{\Delta}

\newcommand{\cF}{{\cal F}}
\newcommand{\cM}{{\cal M}}
\newcommand{\ra}{\rangle}
\newcommand{\la}{\langle}
\newcommand{\tb}{\tilde{\beta}}
\newcommand{\td}{two-dimensional}

\def\void{}
\def\labelmark{}

\newenvironment{formula}[1]{\def\labelname{#1}
\ifx\void\labelname\def\junk{\begin{displaymath}}
\else\def\junk{\begin{equation}\label{\labelname}}\fi\junk}%
{\ifx\void\labelname\def\junk{\end{displaymath}}
\else\def\junk{\end{equation}}\fi\junk\labelmark\def\labelname{}}

{\ifx\void\labelname\def\junk{\end{array}\end{displaymath}}
\else\def\junk{\end{array}\right.\end{equation}}
\fi\junk\labelmark\def\labelname{}\def\junk{}
}

\newcommand{\beq}{\begin{formula}}
\newcommand{\eeq}{\end{formula}}
\newcommand{\beqv}{\begin{formula}{}}

\begin{document}
\topmargin 0pt
\oddsidemargin 5mm
\headheight 0pt
\headsep 0pt
\topskip 9mm

\hfill    NBI-HE-92-35

\hfill August 1992
\begin{center}

{\large \bf

Matter fields with $c > 1$ coupled to $2d$ gravity}
\footnote{Supported by a Nato science collaboration grant.}

\vspace{24pt}

{\sl Jan Ambj\o rn}

\vspace{6pt}

 The Niels Bohr Institute\\
Blegdamsvej 17, DK-2100 Copenhagen \O , Denmark\\

\vspace{12pt}

{\sl Bergfinnur Durhuus}

\vspace{6pt}

 Matematics Institute \\
 Universitetsparken 5, DK-2100 Copenhagen \O , Denmark\\

\vspace{12pt}

{\sl Th\'{o}rdur J\'{o}nsson}

\vspace{6pt}

Science Institute, University of Iceland \\
Dunhaga 3, 107 Reykjavik, Iceland

\vspace{12pt}

{\sl Gudmar Thorleifsson}

\vspace{6pt}

 The Niels Bohr Institute\\
Blegdamsvej 17, DK-2100 Copenhagen \O , Denmark\\

\end{center}

\vspace{12pt}

\begin{center}
{\bf Abstract}
\end{center}

\vspace{6pt}

\noindent
We solve a class of branched polymer models coupled to spin systems
and show that they have no phase transition and are either always
magnetized or never magnetized depending on the branching weights.
By comparing these results with numerical simulations of two-dimensional
quantum gravity coupled to matter fields with central charge $c$ we
provide evidence that for $c$ sufficiently large ($c\geq 12$) these
models are effectively described by branched polymers.  Moreover,
the numerical results indicate a remarkable universality in the
influence on the geometry of surfaces due to the interaction with
matter.  For spin systems this influence only depends on the total
central charge.
\newpage
\addtolength{\baselineskip}{0.20\baselineskip}

\newsection{Introduction}

In the last couple of years there has been a significant progress
in our understanding of two-dimensional gravity coupled to matter
with central charge $c \leq 1$. The situation for $c > 1$ is unclear.
The KPZ formula leads to complex critical exponents
for $1 < c < 25$ and the continuum formalism indicates
the existence of tachyons for $c$ in this region. On the other hand
the discrete regularization of two-dimensional gravity coupled
to matter \cite{adf,david,kkm} is well defined for any $c$.
The discretized gravity models are
(formally) reflection positive
\footnote{The principle of reflection positivity can be extended from Euclidean
field theory to random surfaces \cite{dfj1}. It is valid for hypercubic random
surface theory, and we expect it to be satisfied in the scaling limit
for triangulated random surfaces embedded in $d$ dimensional space.}
and consequently
they cannot have tachyonic excitations.  A folkelore has evolved to the effect
that the existence of tachyons in the spectrum of the continuum
theory is reflected in dominance of branched polymer-like surfaces in
the discretized models.  It is possible to prove
that the string tension in a class of discrete models of strings
does not scale at the critical point \cite{dfj,ad}.
The fact that the string-tension does not scale and the consistent
assumption of a scaling of the lowest mass excitation associated with
the two-point
function in target space, lead to the branched polymer picture: The surface
of minimal area (depending on the boundary conditions) totally dominates
in the scaling limit, and the only excitations allowed are spikes of
essentially no area, branching out from this surface. Stated slightly
differently, one can say that in these models the entropy of such surfaces
completely dominates the critical behaviour for $d=2,3,\ldots$.

While the above scenario is plausible, one should keep in mind that no
rigorous proof of the branched polymer picture has been given
in the case of dynamically triangulated surfaces (see, however,
\cite{frohlich}).   Extensive numerical
simulations have not provided a convincing support to the
dominance of branched
polymers for a small number of dimensions $d$ whereas for $d$ large
($ \geq 12$) there seems to be a reasonable agreement between the
numerical results and the branched polymer picture.

The question we want to address in this article is that
of universality for $c > 1$, i.e. do the critical properties
only depend on the total central charge of the matter fields?
Interesting observations in this
direction can be found in \cite{bj,bh,ckr}. In \cite{bj,ckr} Monte Carlo
simulations were used to investigate various properties
of multiple $q$-state Potts models coupled to two-dimensional
quantum gravity, and an apparent universality,
characterized by the total central charge $c$ of the system,
was observed.
One surprising aspect was that the universality seemed to apply
to quantities which one does not expect to
be universal, like the average number of vertices
of order three in the triangulations, calculated at the critical point
of the multiple Potts model. In this paper we extend the analysis
of \cite{bj,ckr} and verify that
that the  distributions of orders of vertices seem to fall into classes
characterized by the total central charge of the system.
In \cite{bh} a strong coupling expansion is used for the calculation
of $\g_{string}$ for  various combinations of Ising spins and of
Ising spins and gaussian fields. Again universality is observed within
the accuracy of the strong coupling expansion.

We first consider multiple spin models on random trees and verify that
they have no phase transition at finite temperature.  We show by an
explicit calculation that if the branching weights are sufficiently
large, so that $\gamma _{string}<0$, then a single Ising model on
a branched polymer is in the magnetized phase at all finite
temperatures.  We emphasize that there is no choice of branching weights
which leads to a finite temperature phase transition for spin systems on
branched polymers.  We then turn to a detailed discussion of the
numerical results.

In summary, we show that there are good reasons to believe
that the branched polymer picture (with some modifications to be discussed
later) is true for sufficiently large values of $c$.
We consider it unlikely to be true all the way down to
$c=1$ since a few spin systems couple almost as a single spin system.

\newsection{Multiple systems coupled to gravity}

In the case of a fixed triangulation (or any other fixed lattice)
the consideration  of $n$ independent copies of a spin system adds nothing
interesting to the information already available in the
partition function $Z(\b)$ of a single spin system. The reason
is that the partition function $Z_n(\b)$ for the multiple system
factorizes:
\beq{*1}
Z(\b)= \sum_{\{ \sg_i \} } e^{\b\sum_{<i,j>} \sg_i \sg_j},
{}~~~~~Z_n(\b) = Z^n(\b)
\eeq
and the free energy $F(\b)=-\ln Z(\b)$, which determines the critical
behaviour of the system  will only differ by a factor from that of a
single spin system.

When we couple the systems to \td\ quantum gravity the situation changes
due to the back-reaction of matter on gravity. In the dynamical triangulation
approach to quantum gravity one replaces the fixed lattice
with a sum over all random lattices (triangulations $T$) with weight
$Z_T^n(\b)$:
\beq{*2}
Z_n(\m,\b)= \sum_T e^{-\m N_T} \; Z^n_T(\b).
\eeq
In this formula $\m$ denotes the bare cosmological constant, while
$N_T$ denotes the number of triangles in the triangulation $T$. We
assume that the topology of triangulations is that of $S^2$.
When we discuss the effect of  multiple spins on  gravity
we see that all they do is to change
the weight of the triangulation from one to
$Z^n_T(\b)$. In general this observation
is not of much help, since it is difficult
to calculate $Z_T (\b)$ for an arbitrary triangulation $T$ and evaluating
$Z_T(\b)$ is not sufficient to judge its significance.
We have to know the entropy factor associated with $T$ and nearby related
triangulations. In the limit where $n \to \infty$ we expect the importance
of the entropy factor to vanish relative to $Z^n_T (\b)$ in accordance
with standard mean field arguments, and for a given $\b$ the triangulations
with the largest partition functions is expected to dominate.

\subsection{Gaussian multiple systems}

The simplest systems with central charge $c\geq 1$ are multiple
Gaussian systems. If we couple them to \td\ gravity
we get Polyakovs formulation
of string theory. The discretized version takes the form
\cite{adf,david,kkm}
\beq{*4}
Z(\m,\tb)= \sum_{T} \rho (T) e^{-\m N_T} \int \prod_{i\in T\setminus
\{ i_0 \}} dx_i
e^{-\tb \sum_{<i,j>} (x_i-x_j)^2}
\eeq
where $N_T$ denote the number of triangles and $i_0$ is a vertex which
is kept fixed in order to eliminate the translational mode. $Z(\m,\tb)$
is independent of the choice of $i_0$.  The function
$\rho(T)$ denotes a weight assigned to each
triangulation $T$. In the above discussion of spin systems we tacitly
assumed that $\rho(T)=1$ for all $T$. This choice is the one which allows
for an explicit solution in the case of the Ising model. Other weights
have been considered in the past. One choice, which seems particularly well
motivated assigns the weight
\beq{*5}
\rho(T) = \prod_{i} n_i^{-\a}
\eeq
to the triangulation $T$. Here $n_i$ denotes the order of vertex $i$,
i.e. the number of links to which $i$ belongs. The case
$\a=0$ corresponds to $\rho(T)=1$,
while $\a = -c/2$ corresponds to the so-called conformal weight
for the path integral in quantum gravity. Since \rf{*4} is a discretized
version of (\td ) quantum gravity, which is not conformally invariant,
there is no obvious reason for choosing one weight rather than another, as
long as the weights satisfy a number of general requirements outlined
in \cite{adf}. By universality the detailed choice of $\rho$ should not
be important.

One difference between the Gaussian system and the discrete spin systems
we are going to consider is that the coupling constant $\tb$ in \rf{*4}
 can be absorbed in a
redefinition of the cosmological coupling constant $\m$
by scaling the Gaussian variables $x_i$.
This reflects that the Gaussian field on a regular lattice is automatically
critical as soon as the infinite volume limit is taken. In the rest
of this article we assume that such a rescaling has been performed
so that $\tb =1$. The coupling constant $\b$ will be used only
for the discrete spin systems.

Another difference between the Gaussian system and the discrete spin
systems is that we can find the part of the partition function corresponding
to a specific triangulation explicitly:
\beq{*6}
Z_T = \int \prod_{i\in T/\{ i_0 \}} dx_i
e^{- \sum_{<i,j>} (x_i-x_j)^2} = {\pi}^{(V_T-1)/2}
 \left(\det( C_T(i_0))\right)^{-c/2}
\eeq
where $V_T$ is the number of vertices in $T$ and
$C_T(i_0)$ is the so-called incidence matrix of the triangulation. Its
entries are labelled by pairs of vertices from $T\setminus\{i_0\}$,
and equal -1 if $i$ and
$j$ are neighbours, $n_i$ (the order of vertex $i$) if $i=j$ and 0 otherwise.
Again $\det (C_T (i_0))$ is independent of $i_0$ and we suppress the
index $i_0$. The partition function \rf{*4} can now be written
\beq{*7}
Z(\m) = \sum_{T} \rho(T) \left(\det (C_T)\right)^{-c/2}
\; e^{-\m N_T},
\eeq
with a redefinition of $\m$.
Much is known about the function $\det C_T$ and  this allowed
the determination of
the class of triangulations which dominate the partition function  for
large $d$ \cite{adfo,adf1,bkkm}. One has to go to large $d$ in order
to be able to apply mean field theory, i.e. in order to be able to ignore
the entropy factors associated with triangulations.
A class of triangulations which leads to an explicit solvable
model and which includes those with minimal determinants for a
given number of triangles $N_T$ was found in \cite{adfo,adf1}. The class
is constructed by gluing together tetrahedra which are cut open along
two non-adjacent links. There is a natural correspondence
between surfaces of this class and branched polymers, such that each
tetrahedron corresponds to a link in the branched  polymer. Along
the links cut open it is possible to glue many tetrahedra, each of
which corresponds to a new branch in the associated polymer. In this
way the order of the branching is not restricted. The determinant
$\det C_T$ can be calculated for this class of triangulations and one
finds:
\beq{*8}
{\det C_T = \prod_{i \in T} n_i , }
\eeq
where $n_i$ is the order of vertex $i$. We see that $\det C_T$ for this
class of triangulations is of the same form as weight factors
$\rho(T)$ considered in \rf{*5}. It is now possible to
rewrite the partition function for $d$ Gaussian fields on the
class of random surfaces described above as
\beq{*9}
Z(\m)= \sum_{BP}  \rho (BP) e^{-\m N_{BP}}
\eeq
where the sum is over all branched polymers,
$N_{BP}$ is the number of links in the branched polymers and the weight
function $\rho (BP)$ according to \rf{*5} and \rf{*8} is given by:
\beq{*10}
\rho(BP) = \prod_{i \in BP} w(n_i),~~~~~ w(n) = \frac{1}{n^{c/2+2\a}}.
\eeq
The ``cosmological constant'' $\m$ is trivially related to
the original $\m$ for the surfaces by a linear transformation, since various
integration factors have been absorbed in $\m$ for convenience. We refer
to \cite{adfo} for details.

The theory of branched polymers was solved in \cite{adfo} for arbitrary
positive weights $w(n)$, and for arbitrary weights in \cite{adj}.
In the following we will only be interested in the situation where the
weights are positive (the non-positive weights lead to a sequence of
of multi-critical models, by the mechanism which gave rise to the
multi-critical models when discretized random surfaces were allowed to
have negative weights). Let us briefly, for future reference,
review how to solve for the critical exponents (\cite{adfo,adj}).
The self-consistent equation for the partition function for  rooted
branched polymers is shown in fig. 1 and reads
\beq{*11}
Z(\m) = e^{-\m} \left(1+\sum_{n}w(n)Z^n(\m)\right).
\eeq
For future convenience let us introduce the two functions
$f$ and $\cF$ defined by
\beq{*11a}
f(z) = \sum_n w(n)z^n~~~~{\rm and}~~~~\cF(z) = \frac{1+f(z)}{z}.
\eeq
The functional relation between $\m$ and $Z(\m)$ is
\beq{*12}
e^\m = \cF (Z).
\eeq
This equation allows us to determine $Z$ as a function of $\m$ near the
critical point $\m_c$ which is the smallest value of $\m$ for which
\rf{*12} has a solution. Two situations can arise. Either $\cF(z)$
has a local minimum $z_c$, which is necessarily quadratic since all
weights are positive,
or the series which defines $f(z)$ has a finite
radius of convergence $z_c$ and $\cF'(z_c) < 0$ exists, in which case
$\cF(z_c)$ is still the absolute minimum.
The critical behaviour is determined by  expanding $\cF$ to the left of
$z_c$. In the first case we get in the neighbourhood of the critical point
$\m_c$:
\beq{*13}
Z(\m)\approx z_c-\frac{2}{{\cF}''(z_c)}\sqrt{\m -\m_c}.
\eeq
This leads to the well known result that the susceptibility exponent
$\g_{string} = 1/2$ in the case where branched polymers of this
kind can be considered a good approximation.

In the other case we have
\beq{*14}
\m-\m_c \approx c_1 (z_c -Z) + c_2 (z_c-Z)^2 + \cdots +c_n (z_c -Z)^{n+\l}
\eeq
where $n+\l$ is the first non-analytic power in the expansion,
$0 < \lambda < 1$.  By assumption
we have $n \geq 1$, and by inverting the above relation we find:
\beq{*15}
Z\approx z_c + d_1 (\m-\m_c)+d_2 (\m -\m_c)^2 +\cdots +d_n (\m-\m_c)^{n+\l}
\eeq
which leads to $\g_{string}= -n+1-\l < 0$.

In the case where $w(n)=n^{-\kp}$ we have  $\g_{string}= -\kp+2$ for
$\kp > 2$.  The two cases ($\g_{string} =1/2$ and $\g_{string} < 0$)
are characterized by different patterns of branching when
the number of links or vertices goes to infinity.
For $\kp < 2$ the model is the ``ordinary'' phase with $\g_{string} =1/2$
and in which vertices of very high order are not too dominant.
For $\kp \to -\infty$ branching is increasingly suppressed, but for finite
$\kp$ the model alway stays in the same universality class. For $\kp >2$
the ratio between the vertices of order one (ends) and the total number
of vertices is large and goes to 1 for $\kp \to \infty$. This means that
there will be vertices of very large order present.
Although this class of branched polymers
might seem rather artificial, it is nevertheless precisely this class
which will dominate if the weight of each triangulation is taken  to be
one and $d$ is large. If we on the other hand choose the so-called
``conformal'' weight $\rho(T)= \prod_i n_i^{c/2}$,
we stay in the ``ordinary''
branched  polymer phase for large $d$. This analysis was originally
performed
in \cite{adf1} and is confirmed by numerical simulations.

\subsection{Branched polymers coupled to Ising  spins}

It is most natural to couple Ising spins to branched polymers
by associating the spin variables with the vertices of the
branched polymers. If one wants to go back to the random surface
picture and consider the polymers as coming from successive gluing
of tetrahedra the spin assignment is slightly more complicated, but
by universality the critical properties should be the same. The partition
function can be written as:
\bea
Z(\m,\b,h)&=& \sum_{BP} e^{-\m N_{BP}} \rho(BP) Z_{BP}(\b,h) \label{*25} \\
Z_{BP}(\b,h) & =& \sum_{\{ \sg_i\}}
e^{\b \sum_{<i,j>} \sg_i\sg_j +h \sum_i \sg_i} \label{*26}
\eea
where the summation $\sum_{\{ \sg_i\}}$ is over all Ising spin configurations
on the branched polymer, while $\sum_{<i,j>}$ is a sum over
all neighbour pairs
of vertices in a given branched polymer. Finally $h$ denotes an external
magnetic field.

Let us denote by $Z_{+}(\m,\b,h)$ the partition function for rooted
branched polymers with the spin fixed to be up at the root and let
$Z_{-}(\m,\b,h)$ be the partition function with the spin down at the root.
If we use the convention that no magnetic
field is attached to the root we get the following equations:
\bea
Z_+ & = & e^{-\m} \left[2\cosh (\b+h)+
e^{\b+h} f(Z_+) + e^{-(\b+h)} f(Z_-) \right] \label{*27} \\
Z_- & = & e^{-\m} \left[ 2\cosh (\b-h)+
e^{-(\b-h)} f(Z_+) + e^{\b-h} f(Z_-) \right]
\label{*28}
\eea
In the case where there is no external field we have by symmetry $Z_+ = Z_-$
and the solution to \rf{*27} and \rf{*28}
is identical to the solution without Ising spins attached to the polymers
except for a displacement of the critical point
\beq{*29}
\m_c(\b)=\m_c(0)-\ln (2\cosh \b).
\eeq
Until now we have considered the partition function $Z(\m)$ where the
``volume'' $N$ is allowed to fluctuate (the grand canonical
partition function).
It is related to the canonical partition function $Z_N$ by
\beq{*30}
Z(\m) = \sum_N Z_N e^{-\m N}
\eeq
and the asymptotic behaviour of $ Z_N$ for large $N$ is
\beq{*31}
Z_N = N^{\g-2}\; e^{\m_c N}\;\left(1+O(1/N)\right).
\eeq
Since the bulk properties of the spin system are determined by the
the infinite ``volume'' limit of the free energy $F(\b)$
we get
\beq{*32}
F(\b) = \lim_{N \to \infty} \frac{ \ln Z_N (\b)}{N} = \m_c(\b).
\eeq
The infinite ``volume'' limit of the free energy of the spin system is
just the critical value $\m _c(\b)$ and {\it we see from \rf{*29} that
the system of branched polymers never shows any critical behaviour
when coupled to spin}.  The dependence is the same as in a linear
chain of spins, a fact which reflects the {\it tree-structure} of
the branched polymers we consider here.   {\it This result can easily
be generalized to multiple Ising models coupled to branched polymers.}

\vspace{12pt}

\noindent
Although there is no critical behaviour for Ising spins coupled to
branched polymers their behaviour is not necessarily identical to
that of a linear chain of spins. In the last subsection we saw that
there (in the case of positive weights)  existed two
classes of branched polymers characterized by
$\g = 1/2$ and $\g < 0$, respectively.  The latter class had a
branching pattern dominated by a few vertices of very high order and it
has an infinite
Hausdorff dimension, both intrinsically and embedded in target space
in agreement with the results of \cite{tj}, while
the $\g=1/2$ branched polymers have intrinsic Hausdorff dimension two
\cite{adj} and
Hausdorff dimension four when embedded in target space. We will now show
that the ``ordinary'' branched polymers with $\g=1/2$
allow no spontaneous magnetization,
in accordance with their tree-like structure which is much like a linear
chain, while Ising spins on branched polymers
with $\g < 0$ are always  magnetized. The fact that
the spins couple more strongly for large
$d_H$ is in accordance with intuition, but {\it the fact that
ordinary branched polymers with intrinsic Hausdorff dimension
2 have no phase transition
shows that the intrinsic Hausdorff dimension is not always a reliable guide
to the critical properties of systems with fractional dimension
like the ones considered in this article}.

In order to prove the above statements we expand equations
\rf{*27} and \rf{*28} to first order in $h$ to determine
the change $\Delta \m(\b,h)$ of the critical point due to a
small change in $h$.
According to \rf{*32} this allows us to
calculate the spontaneous magnetization:
\beq{*33}
\la M \ra = \lim_{h \to 0^+}
 \frac{\partial F (\b,h)}{\partial h}.
\eeq

We treat the cases $\g=1/2$ and $\g < 0$ separately. For $\g=1/2$
we get (after some algebra) by expanding around
the critical point \rf{*29} and using the notation
$Z_{\pm} (h,\b)= z_c(\b) +\Dl Z_{\pm}$:
\beq{*34}
 (\Dl Z_+ + \Dl Z_-)\left( e^{\m_c(\b)}-f'(z_c)\right)+
2 z_c e^{\m_c (\b)} \Dl \m =0
\eeq
\beq{*35}
(\Dl Z_+ -\Dl Z_-)\left( e^{\m_c(\b)}-\tanh (\b) f'(z_c)\right) =
4h(1+f(z_c))\; \sinh(\b) .
\eeq
Since $ e^{\m_c(\b)}=\cF (z_c(\b))$ where $z_c(\b)$ is determined by the
requirement that
\beq{*35a}
\cF' (z_c(\b)) = 0~~~~{\rm or} ~~~~ f'(z_c(\b)) = \cF (z_c(\b))= e^{\m_c(\b)}
\eeq
 we conclude that
\beq{*36}
\Dl \m = O ( h^2),~~~~~~ \Dl Z_+ - \Dl Z_- =
h \left(e^{2\b} -1\right) z_c(\b).
\eeq

In case where $\g < 0$~, $z_c (\b)$ coincides with the radius of convergence
of $\sum_n w(n) z^n $ and $\Dl Z_+=0$ (since it is $\geq 0$ if we assume
$h\geq 0$).
If $z_c$ denotes
the radius of convergence, the linearized
equations for $\Dl Z_-$ and $\Dl \m$ can now be solved and we get:
\beq{*37}
\Dl \m(\beta ,h) = h \tanh (\b)
\left[ \frac{\cF'(z_c)}{\cF'(z_c) + (-1+\tanh (\b))f'(z_c)/z_c} \right].
\eeq
Since $\cF' (z_c) \leq 0$ and $f'(z_c) > 0$ for $\g < 0$
the above expression is well defined  and the system is magnetized
for $\b > 0$ except in the borderline case $\cF '(z_c)=0$.

\subsection{Multiple Ising spins coupled to \td\ gravity}

The Ising model coupled to \td\ gravity was solved in the seminal papers
\cite{kazakov,bk}. Unfortunately it seems impossible
to use the same analytic technique for multiple
Ising models coupled to gravity. This is why we in the next sections
will turn to numerical methods. It is, however, possible to give a few
arguments which show how difficult it is to judge what will be the
effective back-reaction of multiple spin systems. One could be tempted
to argue as follows: Since the Ising model on a regular lattice, at
the critical point, has a fermionic representation, it is
natural to expect that one can just replace the bosonic determinant
term $(\det C_T)^{-c/2}$ in \rf{*7} by $(\det C_T)^{n/2}$, where $n$ is the
number of Ising spins. The effect of this would be drastic for large
$n$ since it would be equivalent to taking $c \to -\infty$. This $c$-limit
is known to correspond to dominance of regular or quasi-regular
triangulations (\cite{adfo}), quite contrary to the dominance
of branched polymers for $c \to \infty$. Thanks to the exact solution
of the Ising model coupled to \td\ gravity we know the above argument is
incorrect for a single spin and the numerical results presented in the
next section do not support it for multiple spin systems either.

\newsection{Summary of results}

Inspired by the observations in \cite{bj} we have coupled
multiple $q$-state Potts models to \td\ quantum gravity for $q=2,3,4$.
These are the simplest discrete systems which on a regular lattice
exhibit a second order phase transition and to which we can associate
a central charge $c_q$. To be precise we place the spins at
the (centers of) the triangles in the (abstract) triangulation
and let them interact with the spins on neighbouring triangles. This
is equivalent to placing the spins at
the vertices in the dual $\phi^3$ graph.

The coupling to gravity will influence
the critical properties of the system and the back-reaction of the
spin system will modify the critical properties of \td\ gravity.
For a single spin system the back-reaction is in a sense weak, since
it is known that the critical
properties of \td\ gravity are only changed {\it at} the critical
point of the spin model. Whether the coupling remains weak
in this sense for many spin systems coupled to gravity is unknown, but as
we shall see the numerical results indicate that this is not the case.

We have verified numerically
the existence of a transition between a magnetized phase
and one where the average magnetization is zero, when many spin systems
are coupled to two dimensional gravity. This transition seems
to change its character when the number of systems increases
 as we describe below. From general considerations
it appears most likely that there will always be a transition, no matter
how many spin systems we couple to gravity. The reason is the following:
For sufficiently small $\b$ (high temperature)
a single spin system is disordered for a whole
range of temperatures and effectively the coupling of spins to gravity
disappears when $\b \to 0$. In this limit the multiple spins will
consequently be independent. For large $\b$ we have a similar decoupling
between spin degrees of freedom and gravity and we know that for a single
spin system there is spontanous magnetization, which for large $\b$ goes to
one. In this limit there is no interaction between gravity and spin. It is
therefore consistent that the multiple spin system for large $\b$ moves
to a phase with magnetization. However, between the limits of small
and large $\b$ the coupling between gravity and spins can (and will)
be strong and it depends on the number of spin systems
seemingly parametrized to some extent by their central charge.

The back-reaction of the spin systems on gravity
will manifests itself in a change of the average geometrical
properties of the random surfaces. Whether such a change leads
to a different critical behaviour is much more difficult to
decide. We have already mentioned that a single Ising spin only influences
the critical exponents of \td\ gravity when $\b=\b_c$, but it gives of
course  weights $\rho(T) = Z_T(\b) \neq 1$ for each triangulation
when $\b \neq \b_c$.

We have tried to estimate the back-reaction of the multiple spin
system by measuring a number of quantities:
The distribution of vertices of different orders, the intrinsic
Hausdorff dimension and the average number of branches of a surface.
For all these quantities we have compared the results with similar
measurements made on multiple Gaussian systems coupled to gravity.

In order to define concepts like the average radius of a triangulation and
the (intrinsic) Hausdorff dimension of the ensemble of triangulations
we need the concept of geodesic distance on the triangulations.
While the geodesic distance has a unique definition if we view the
triangulation as a piecewise flat manifold glued together of
equilateral triangles with the  curvature assignment of Regge calculus, this
definition is not convenient from a computational point of view. It is more
convenient to define the geodesic distance between two vertices of a
triangulation as the length of (one of) the shortest path within the
triangulation which connects the two vertices. In addition we can talk about
the shortest path between two triangles and define it to be the smallest
number of triangles which constitute a connected path of triangles between
the two. This length between two triangles obviously becomes the the shortest
path between two vertices along links in the dual triangulation.
With these definitions of geodesic length $r$ on a triangulation
we can define the Hausdorff dimension $d_H$ in the usual way:
\beq{*41}
V(r) \propto r^{d_H},
\eeq
where $V(r)$ denotes  the number of triangles within geodesic distance
$r$ of a given vertex or triangle.
An average of $V(r)$
over the given manifold and over the ensemble of manifolds should now
be performed in order to extract $d_H$ from \rf{*41}.
We can also define the (average) maximal
radius $r_{max}$ of two-dimensional (piecewise flat) manifolds in the
following way: Take an arbitrary vertex (the ``center'') and mark it.
Mark all its neighbours. These are the vertices at distance one from the
center. For all these points find their neighhours which are not
already marked. These are the vertices a distance two from the center.
In this way it is easy (from a numerical point of view) to compute
$V(r)$ defined by \rf{*41}. When all points are marked we have the maximal
extension  of the manifold with respect to the chosen center.

\vspace{12pt}

\noindent
Let us now qualitatively summerise our observation and draw some conclusions.
The detailed description of the numerical data will be given in the
next section.

\vspace{12pt}

\noindent
For the magnetic properties of the spin systems we have seen a transition from
a phase with zero magnetization to a phase with magnetization. However
with increased central charge the critical value of $\b$,
$\b_c$, increases. It appears
to be more difficult to magnetize the system. When the transition finally
takes place, the cross over from the non-magnetized phase
to the magnetized phase is
 increasingly rapid. At any given value of $\b$ the magnetization
is a decreasing function of the central charge for a given kind
of spin systems. The phase transition point becomes increasing difficult
to localize by standard methods and the value of the magnetization
{\it at} the critical point increases with increasing $c$. Many aspects
of the systems (specific heat etc.) look for large $c$ quite similar
to the results one obtains by coupling a single spin system {\it and} a
multiple Gaussian system with the same central charge to gravity. One
difference
is that the latter system seems to have no phase transition in the spin
variables.

These properties are what we would expect if we had a
random surface system which degenerated to a
branched polymer system with $\g =1/2$  as
the multiplicity of the spin system increases
\footnote{In order to avoid confusion
the following should be made clear: Since the spins live
on the triangles it is
appropriate to view them as located on the vertices of the dual lattice.
The dominant configurations of branched polymers  generated by the gaussian
action discussed earlier are polymers which in the dual lattice look
like long chains\cite{adfo,adf1}  which should in a natural way be associated
with polymers of $\g_{string}= 1/2$.}. In that case one would
see a decreasing magnetization as a function of multiplicity
as the random surface system changes towards branched polymers
with $\g=1/2$ which have no magnetization, as shown above.
For sufficiently
large $\b$ the coupling will however weaken. The random surfaces
return to their pure gravity phase, independent of the spin systems and
the magnetization of the spin systems will be like that of a single system.
If we however, as mentioned above, couple a single spin system and
multiple Gaussian systems, the spin system is a small perturbation
and the surfaces are forced to stay in the branched polymer phase
where there is no transition. From this point of view there should be agreement
{\it below the critical $\b_c$} between multiple spin systems and multiple
Gaussian plus one spin system. And this seems qualitatively to be the case.

The above observations raise the question at what temperature we should
compare the spin systems to a Gaussian system.
For large central charge the back-reaction from the spin
systems on the geometry seems to be largest {\it before we reach $\b_c$}.
This is seen very clearly when we measure $r_{max}$, the maximal radius
of the surfaces, as defined above. There is an
extended region before $\b_c$ where the influence of the spin systems on
the geometry seems to consistent with, and at least as strong as that of
a similar multiple Gaussian system. Intuitively this is in agreement
with the hypothesis that surfaces in this region degenerate into
 branched polymers: The strong influence
of the spin systems deforms the geometry to that of branched polymers and
prevents thereby their own criticality since, as we have shown above, there
can be no critical behaviour for branched polymers. It is only when $\b$
becomes
sufficiently large that the spins decouple (close to $\b_c$) and
the geometry returns to that of \td\ gravity.

{}From this point of view the situation is
very different from the case of  a single Ising system
where we know that only {\it at} the critical
point is the coupling between spin and gravity sufficiently strong to
change $\g_{string}$. We conjecture that for the central
charge $c$ sufficiently large ($ c \geq 10-12$) there is a whole
range of $\b$ ending at $\b_c$ where the geometry has undergone a
transition to branched polymers like the ones induced by
a multiple Gaussian system with the same central charge.

\newsection{Data}

The numerical results reported in the following sections is based on
micro-canonical Monte Carlo simulations where lattice sizes ranged
from 500 to 8000 triangles $N_T$. The standard ``link flip''
algorithm \cite{kkm} was used to update the random triangulation,
while the Swendsen-Wang cluster algorithm \cite{sw}
was used to update the spins.
This algorithm is known to be very efficient in eliminating critical
slowing down close the spin phase transition.

Usually 40.000-100.000 sweeps were used, where one sweep consisted
of updating the whole spin configuration and of performing
$N_T$  link-flips.

As alreay mentioned we also looked at systems with
Gaussian fields coupled to two-dimensional gravity. The Gaussian fields
were placed on the triangles in analogue with the spin systems and
a simple Metropolis algorithm was used to update them.

The measurements fall in two classes:
Measurements of the ``magnetic'' properties of the spin systems
when they are coupled to \td\ gravity and measurements of quantities
which characterize the intrinsic geometry of the triangulations,
and thereby the back-reaction of the spins on \td\ gravity.

\subsection{Magnetic properties of the spin systems}

\subsubsection{Multiple Ising spins}

We have looked at 1,2,4,8 and 16 Ising spin models coupled to gravity.
In fig. 2 we
have shown  typical magnetization curves for one Ising system
and sixteen
Ising systems coupled to \td\ gravity. The value of the
average magnetization $|\cM |$ per spin per volume decreases with
the multiplicity of the spins and the transition to a magnetized state
is pushed to larger $\b$.

In fig. 3 we show Binder's cumulant
\beq{*40}
BC= \frac{ \la M^4 \ra}{(\la M^2 \ra)^2}-3
\eeq
for the two systems. It is well known that the use of Binder's cumulant
in the case of regular lattices offers an efficient way
to decide whether we have a first or higher order transition and
in addition, if the transition is of higher order, allows us to determine
the critical point. $BC$ as a function of $\b$ is much
smoother for a higher order transition than it is in the case of a
first order transition. In addition $BC$ for different
volumes will intersect in a common point which is the critical point at
infinite volume.  In the case of dynamical triangulations the method works
not quite as well, and one has to go to somewhat larger systems
($N_T > 1000$) than for regular lattices in order
to see a clear picture. For a single or a few spin systems we eventually
see curves corresponding to a higher order transition and curves
of different (sufficiently large) volume intersect in the same point.
As the number of spin systems increases the signal of a phase transition
deteriorates. It becomes more difficult to identify a common intersection
point and it (in case it really exists) moves to larger $\b$-values
and the value of $BC$ at the transition moves closer to $BC=-2$,
which corresponds to magnetization close to one. From
this point of view we are far from the nice situation of a single
Ising spin where the intersection point for $BC$'s corresponds
to $|\cM| \approx 1/2$. We have illustrated this in fig. 4 in a
curve which exhibits the average value $|\cM|$ at the critical point
as a function of the multiplicity  of Ising spins.

In fig. 2 we also show for comparison
the effect of coupling gravity to a multiple Gaussian systems.
In this case it is known that the surfaces degenerate into branched
polymers for sufficiently many gaussian fields.
To this system we have now coupled a single Ising spin. The magnetization
curve looks much like the one for a branched polymer without constant
magnetization or a one-dimensional chain. This similarity becomes more
pronounced as the multiplicity of the Gaussian system increased from
8 to 16. Effectively one sees a
magnetization  $|\cM | $ for a finite value, but the slope is soft and
gets softer with increasing volume. In addition it is impossible to
find an intersection point for $BC$ (fig. 5).
The main difference
compared to the case of multiple Ising spins is that we have
no phase transition to a magnetized state  since  the single spin
system is only a small perturbation of the gaussian fields which, by their
back reaction, force the geometry of the triangulations to be that of
branched polymers. In fact the results of computer simulations of a single
Ising spin directly on branched polymers look similar to  the ones
shown here.

\subsubsection{Multiple q-state Potts models}

For $q$-state Potts models ($q=3,4$) we have performed similar
measurements to those reported above for the Ising model, which corresponds
to $q=2$. We have coupled 1,2,4,5,8,10 and 16 $3$-state Potts models
to two-dimensional gravity and the same has been done for 1,2,4,8 and 16
4-state Potts models. The results for the magnetic properties are in
agreement with those found for the multiple Ising models, i.e. the
magnitization decreases with the multiplicity of the spin systems and the
critical points, found as the intersection of the $BC$-curves, are located
at a value of $BC$ closer and closer to -2.

The critical points change with the multiplicity, reflecting the interaction
between the different spin systems  mediated by  two-dimensional gravity.
One surprising aspect is that the displacement of the critical point
seems to be a simple (almost linear) function of the multiplicity, identical
for $q=2,3$ and 4. This is in contrast to the coupling to gravity which
as we shall see later can be grouped according to the central charge $c$
of the multiple spin system.

\subsubsection{Specific heat and susceptibility}

The direct measurements of the magnetic susceptibility did not reveal
any signals which were not compatible with it being the derivative of the
magnetization curve. The peak moved to somewhat larger $\b$ values
and got steeper with  increasing multiplicity.

The specific heat peak does not scale with lattice
size for any of the systems. The peak gets less pronounced
with increasing multiplicity. Again this happens relative
smoothly as a function of multiplicity. When we compare the curves
for sixteen Ising models coupled to gravity to that of
one Ising model and eight Gaussian models coupled to gravity they look
almost identical and qualitatively quite similar to the curves one would get
for a linear spin chain (or a branched polymer chain)
where we know there is no transition, but where it is trivial to show that
the specific heat nevertheless has a peak for a finite $\b$
like the one shown in fig. 6
for sixteen  Ising models and for one Ising model plus
eight Gaussian models. For comparison we have shown the
peak for one Ising model coupled to gravity.

 From this observation
it is tempting to guess that this reflects the linear structure
(branched polymer structure) of the \td\ surfaces for such high
value of the central charge.

\subsubsection{Finite size scaling results}

We have applied the methods of finite size scaling to the data obtained
at the critical points $\b_c$ for the various models in order to extract
the critical behaviour of observables associated with the magnetic
transition. In order to follow the usual philosophy of finite size
scaling we must first extract a linear scale $L$ for the system.
The only natural choice is
\beq{*2.30}
L \sim N_T^{1/d_H}
\eeq
where $N_T$ is the number of triangles of the system and $d_H$ denotes
the Hausdorff dimension. Finite size scaling then leads to the
following relations at the critical point for the magnetization
$\cM$, the magnetic susceptibility $\chi$ and the specific heat $C_V$:
\beq{*2.31}
\cM \sim N_T^{-\b/\n d_H},~~~~\chi \sim  N_T^{\g/\n d_H},~~~~
C_V = B + C_0 N_T^{\a/\n d_H}.
\eeq

We have used lattices with $N_T=500-4000$ and performed measurements
for all spin models we have examined. For the magnetization and the
susceptibility we get good linear fits to the log-log plots and the
results are presented in table 1. We see reasonable (although
not impressive) universality  with respect to central charge.
The trends are the same for all spins: $\g/\n d_H$ increases with
$c$ while $\b/\n d_H$ decreases.

It is more problematic to extract the specific heat exponents due
to the constant term $B$ in \rf{*2.31}. The fits we made indicated
that $\a/\n d_H$ approaches zero as the multiplicity is increased, but
we do not consider this result reliable.

Let us end this section with a few remarks about the use of finite size
scaling in quantum gravity. The standard arguments rely on the concept
of a linear scale which characterizes the system of a given volume.
In case there exists a Hausdorff dimension the identification
\rf{*2.30} is sensible, but it is not clear whether there
is any well defined Hausdorff dimension in \td\ quantum gravity.
There are many indications that $d_H$ should be considered infinite
\cite{tj,am}. It is hard to believe that the interaction
of a single Ising system with gravity should change this. In the same
way it is hard to believe that the concept of a correlation length
which scales makes strict sense in such a space.
Probably one should view $\n$ as zero in this context and since
the finite size scaling seems  to give reasonable results it
might be that the product $\n d_H$ is well defined. It would
be quite interesting to understand this better in case it is
proven that $d_H = \infty$.

\subsection{Back-reaction on gravity}

We now turn to the measurements of quantities related to the geometry
of the random surfaces.

\subsubsection{The maximal radius}

The simplest quantity to measure  is the maximal  radius $r_{max}$ defined
above. It turns out to be the most interesting, too. In fig. 7
we have plotted $r_{max}(\b)$ for various multiple spin systems as a
function of the coupling constant $\b$. We see precisely the picture
advocated during the discussion in the last section.
For all spin systems $r_{max}$ decreases as
$\b$ increases from zero.
This we take as a sign of an increased coupling between
the dynamical triangulation and the spin systems. Close to the magnetic
phase transition $r_{max}(\b)$ rapidly increases to its value without
spins: The spin systems are magnetized and effectively decouple from
gravity. However, the important point is that $r_{max}$ seems to
be lowered quite significantly with increasing multiplicity of the
spin systems. Although there is not really any stringent universality
with respect to central charge for these curves we can say that
when the total central charge gets larger than 6-8, then $r_{max}$ seems
to dive to its smallest values {\it before} $\b$ reaches
its critical value $\b_c$. In the region where the
system magnetizes there is an increasingly  rapid
change back to the size of random surfaces without spin systems.
It is tempting to view the transition in geometry as one
between a branched polymer phase (corresponding to that induced by the
multiple Gaussian interaction) for $\b$ in the $\b$-region
of low $r_{max}(\b)$  to
a phase where the geometry is like in the absence of spin systems.
For comparison we show in fig. 8
a graph of $r_{max}$ as a function of $c$ for
multiple Gaussian systems where there also is a rapid decrease of
$r_{max}$ with increasing $c$.

\subsubsection{Fractal properties}

Another obvious geometrical quantity to measure is the Hausdorff dimension.
We have tried to extract it. As known it is quite problematic even
in pure \td\ gravity \cite{am}, and this is unchanged
when gravity is coupled to multiple spin systems. One observes the same
phenomenon as for pure gravity, namely that the Hausdorff dimension seems
to drift towards higher values with the system size.
This drift is seen for both definitions of
the geodesic distance given above, but most clearly in the case where
the ``triangle distance'' was used. Consequently it is difficult to extract a
unique Hausdorff dimension and it makes better sense to compare
the Hausdorff dimension for surfaces with different spin systems in order to
reveal the general trend in change of geometry as the multiplicity of
spins is varied.
In fig. 9 we show the curves of radius versus volume for different
spin systems, again at their tentative critical point. We observe
universality with respect to central charge of the multiple spin
systems, but no precise agreement if we compare with  multiple gaussian
systems of similar similar central charge.
The fact that there is not agreement with the Gaussian systems for large
values of $c$ is actually in agreement with the observation that the
spin systems are increasingly magnetized at $\b_c$ for large $c$ and therefore
couple more weakly to gravity. In fact $r_{max}$ is almost constant
{\it at} $\b_c$ as a function of $c$, even if it decreases drastically
for $\b$ slightly smaller than
$\b_c$ as a function of $c$, cf. fig. 8.
It is
the region below $\b_c$ which shares the characteristica of the multiple
Gaussian systems.

It is important to keep in mind that the
Hausdorff dimension is a very rough measure of
geometrical properties of the system and it might might not tell us
much about the geometry. As examples we mentioned that both smooth surfaces
and branched polymers with $\g =1/2$ have  intrinsic Hausdorff dimension two.
We have therefore attempted to get a more detailed knowledge of the
fractal properties of the geometry by measuring the number of connected
components of the set of points at
a certain distance from a fixed point (or fixed triangle). In the
case of smooth surfaces one would measure few such components, while
in the case of surfaces which resemble branched polymers we expect to
see many components. The results of the measurements at the critical points
are shown in fig. 10.  We see an increase in the  number of
components as a functon of $c$ for the Gaussian systems, but
again the spin systems show a weak coupling to the geometry.
The qualitative trend
is towards many components and therefore increased branching and again
there seems to be universality with respect to the central charge of the
spin systems.

The branching is less pronounced than for Gaussian systems of the
same central charge $c$. If we however go {\it below} the critical point
we see the stronger coupling to gravity already observed for $r_{max}$.
The effect is even slightly larger in a $\b$ region
below $\b_c$  than the corresponding
effect induced by a Gaussian system with the same central charge.

\subsubsection{Distribution of orders of the vertices}

The first observation is a confirmation and extension of the result
found in \cite{bj}: If we consider the distribution of orders of vertices
for multiple $q$-state Potts models ($q=2,3,4$ corresponding to
central charge $c_q=1/2,4/5,1$) at their critical points, {\it then
 combinations with identical central
charge seem to lead to identical distributions}.
This is a strong argument in favor of the
existence of some kind of universality for random surface theories
even if $c > 1$. Of course the distributions are only identical within the
errorbars of the simulations, but the errorbars are quite small and while
(for instance) five $q=4$ state Potts models lead to the same distributions
as 8 Ising models coupled to gravity, there is a marked difference
when we compare with the distribution of five Ising models. This is illustrated
in figs. 11 and 12.  It is interesting to show in addition the fraction
of vertices of order three as a function of $\b$ for the various multiple
spin systems (fig. 13). It again reflects very clearly that the coupling to
geometry is much stronger below $\b_c$ than at the critical point, especially
for multiple spin systems. As expected by now there is not any agreement
if we compare the  distributions of vertices {\it at the critical $\b_c$'s}
with similar distributions in the multiple Gaussian models
of the same central charge. In order to get qualitative agreement one has
to go below the critical $\b$'s.

\newsection{Discussion}

We have tried to check the claim of universality put forward in
\cite{bj,bh,ckr}. We have verified and extended the remarkable universality
in distributions of the orders of vertices of the multiple
$q$-state Potts models as a function of the central charge.
The universality seems also to extend to the fractal properties of
triangulations coupled to multiple spin systems.

However, there is not universality when we allow Gaussian fields
on the surfaces. This does not necessarily imply that the critical
properties of \td\ gravity coupled to non-Gaussian fields are different from
the critical properties of \td\ gravity coupled to Gaussian fields of the
same central charge. There is no reason why
the detailed distributions of the order of the vertices should be the same.
The Hausdorff dimension is related to the critical exponents and should be
the same if  finite.
Qualitatively the coupling of multiple spins
and Gaussian fields to \td\ gravity lead to the same back-reaction on gravity.
{}From the old grand canonical simulations of Gaussian
fields coupled to random surfaces it is known that the
system moves to a branched polymer phase for large values of the central
charge $c$. This value is however not very precisely determined, but
is compatible
with a value where the spin systems start to interact strongly
with gravity, where the radius of extension $r_{max}$ becomes small
and the critical point becomes difficult to identify by means of
Binders cumulant. All this is consistent with a branched polymer
picture, but it is also clear from
the data presented that we cannot claim much more than consistency.
The transition to branched polymers is too soft to give
a clear signal for the size of systems we are able to study by numerical
simulations, or maybe the the transition is simply of a
new kind and we have just not found the right quantity
to charactize it.

One could ask why we did not try to measure $\g_{string}$ directly
for the multiple spin systems,
and in this way tested universality.
The reason is that one would have to use
a grand canonical ensemble and the measurements  of $\g_{string}$ for the
Gaussian models were never  satisfactory \cite{afkp}. Repeating those
measurements would at most have yielded a qualitative agreement
with the Gaussian result, which we have already been able to confirm without
use of the grand canonical ensemble.

\newpage

\begin{center}
{\large \bf Tables}
\end{center}

\begin{itemize}

\item[Table 1] Numerical results for the critical exponents $\b/\n d_h$
and $\g/\n d_h$ for
the different number of Potts models coupled to two-dimensional gravity.
  The values are
obtained using finite size scaling and the errors indicated are those of the
$95\%$ confidence levels.
\end{itemize}

\vspace{12pt}

\begin{tabular}{|c|c|c|c|c|c|c|}     \hline
  &
  \multicolumn{3}{c|}{$\b/\n d_h$} &
  \multicolumn{3}{c|}{$\g/\n d_h$}  \\
$N_{models}$  & $q=2$    & $q=3$      & $q=4$
      & $q=2$    & $q=3$      & $q=4$      \\      \hline
$1$           & 0.146(4)   & 0.093(4)   & 0.111(5)
              & 0.803(10)   & 0.861(20)   & 0.824(16)   \\
$2$           & 0.095(3)   & 0.077(3)   & 0.084(4)
              & 0.888(10)   & 0.856(13)   & 0.860(12)   \\
$4$           & 0.073(3)   & 0.085(3)   & 0.087(4)
              & 0.883(11)   & 0.890(13)   & 0.888(12)   \\
$5$           &            & 0.061(3)   &
              &             & 0.894(13)   &             \\
$8$           & 0.053(3)   & 0.055(3)   & 0.059(3)
              & 0.906(13)   & 0.983(13)   & 0.956(18)   \\
$10$          &            & 0.042(2)   &
              &             & 0.933(15)   &             \\
$16$          & 0.059(2)   & 0.044(2)   & 0.039(2)
              & 1.023(14)   & 0.994(15)   & 0.990(12)   \\   \hline
\end{tabular}

\newpage

\begin{center}
{\large \bf Figure captions}
\end{center}

\begin{itemize}

\item[Fig.1] A graphical representation of the equation which
determines the rooted partition function for branched polymers.

\item[Fig.2] Magnetization curves for one and sixteen Ising models
(1 $q=2$ and 16 $q=2)$
coupled to two-dimensional gravity.
Also shown are the curves for a single Ising model and, respectively,
eight and sixteen Gaussian systems coupled to gravity
(1 $q=2$ \& 8 G and 1 $q=2$ \& 16 G).

\item[Fig.3] Binder's cumulant for one and sixteen Ising models coupled
to gravity. The intersections of curves corresponding to the different
lattice sizes
yield the infinite volume critical points.

\item[Fig.4] Average value of the magnetization $|\cM |$ in the critical
points for 1,2,4,8 and 16 Ising system models coupled to gravity.

\item[Fig.5] Binder's cumulant for a single Ising model and sixteen
Gaussian systems coupled to gravity.  It is not possible to find any
clear point of intersection for the different $BC$ curves.

\item[Fig.6] Specific heat curves for sixteen Ising models and
a single Ising model plus eight Gaussian systems coupled to gravity.
For comparison the specific heat curve for one
Ising model coupled to gravity is also shown.

\item[Fig.7] The maximal radius $r_{max}$ as function of the reduced
''temperature`` $\tau = (\b - \b_c)/\b_c$.  Measurements
are shown for 1,8 and 16 Ising models, and 8 and 16 $q=4$ Potts
models, coupled to two-dimensional gravity.

\item[Fig.8] The changes in $r_{max}$ as a function of $c$ for one to
sixteen Gaussian systems coupled to gravity.  For
comparison are shown the values of $r_{max}$, measured in the critical points,
for all the spin models we studied.

\item[Fig.9] A log-log plot of the radius vs volume, measured
at the critical points, for 8 and 16 $q=2$, 5 and 10 $q=3$ and
4 and 8 $q=4$ Potts models coupled to gravity.  Also shown are the curves
for 4 and 8 Gaussian systems coupled to gravity.

\item[Fig.10] Number of connected components $b$ (branching) as a function
of the distance from a fixed point, for the same systems
as considered in fig. 9.

\item[Fig.11] Fraction of vertices of order three as a function of
the central
charge $c$ for one to sixteen Gaussian systems coupled to gravity.  This is
also shown for all the spin models we studied, calculated at their
critical points.

\item[Fig.12] Fraction of vertices as a function of their order
$n_i$ for 16 $q=2$, 10 $q=3$ and 8 $q=4$ Potts models coupled to
gravity, calculated at $\b_c$.  The corresponding  curve for
8 Gaussian systems is also shown.  In all cases the total central charge is 8.

\item[Fig.13] Fraction of vertices of order three as a function of the
reduced  ''temperature`` $\tau$ for the same systems as in fig. 7.

\end{itemize}

\end{document}